# On the Decrease Rate of the Non-Gaussianness of the Sum of Independent Random Variables

Jacob Binia
New Elective - Engineering Services Ltd
Haifa, Israel
Email: biniaja@netvision.net.il

*Abstract*—Several proofs of the monotonicity of the non-Gaussianness (divergence with respect to a Gaussian random variable with identical second order statistics) of the sum of n independent and identically distributed (i.i.d.) random variables were published. We give an upper bound on the decrease rate of the non-Gaussianness which is proportional to the inverse of n, for large n. The proof is based on the relationship between non-Gaussianness and minimum mean-square error (MMSE) and causal minimum mean-square error (CMMSE) in the time-continuous Gaussian channel.

## I. INTRODUCTION

The non-Gaussianness of a random variable (random vector, random process) $\xi$ is characterized by the divergence $D(\xi) = D(\xi \| \tilde{\xi})$ between $\xi$ and $\tilde{\xi}$, where $\tilde{\xi}$ is Gaussian with the same second order statistics as that of $\xi$. Recently, a simple proof of the monotonicity of the non-Gaussianness of the sum of n independent and identically distributed (i.i.d.) random variables was shown in [1]. In particular, if $X_i, i = 1,...,n$ are i.i.d. then

$$D(\sum_{i=1}^{n} X_i) \leq D(\sum_{i=1}^{n-1} X_i). \qquad (1)$$

Results for a more general setting of non-identically distributed random variables were also given in [1]. Moreover, monotonicity of Fisher information, entropy and other information theoretic magnitudes were also analyzed (e. g. [2]-[4]). However, no estimate of the decrease rate of the non-Gaussianness was given.
In this paper we give an upper bound on the decrease rate of the non-Gaussianness which is proportional to the inverse of $n$, for large $n$. The proof is based on the relationship between divergence and both minimum mean-square error (MMSE) and causal minimum mean-square error (CMMSE), in the time-continuous Gaussian channel.
In section II we state our main result (Theorem 1) and give an example of possible application. The proof of the theorem is given in section III.

## II. THE DECREASE RATE OF THE NON-GAUSSIANNESS OF THE SUM OF IID RANDOM VARIABLES

Let $X_i, i = 1, 2,...$ be independent, identically distributed random variables with finite variance. Without loss of generality we assume all random variables to have zero mean. Denote by $D^{(2)}(0)$ the second derivative of the divergence $D(W + \sqrt{q} X_1)$ at $q = 0$, where $W$ is a zero mean, unit-variance Gaussian random variable, independent of $X_1$. Then the following theorem holds.

**Theorem 1:** For any positive number $Q < \infty$ there exists an integer $N$ such that for all $n \geq N$

$$D(\sum_{i=1}^{n} X_i) \leq \frac{1}{2} D^{(2)}(0) \frac{Q^2}{n} + o(\frac{Q^2}{n}). \qquad (2)$$

In (2) we use the notation $o(v_n) = u_n$ for $\lim_{n \to \infty} [u_n / v_n] = 0$.
The proof of the theorem is given in the next section.

Let us note that choosing an "optimal" $Q$ seems to be not trivial. Small $Q$ leads to a small value of the explicit term in the right hand side of (2). But then $N$ must be big (section III (14)). As a consequence the theorem may hold only for the range of very large $n$s.

The theorem could serve several applications. One of them will be addressed here. It is well known [5], [6] that the capacity of the additive non-Gaussian noise channel increases by the amount of the (divergence) non-Gaussianness of the noise, relative to the capacity of the channel with additive Gaussian noise with the same second order statistics and same signal-to-noise ratio. Suppose that a channel is exposed to an interference which is composed of a normalized sum of $n$ independent interferers (of the type (9) bellow). Then, one can evaluate the capacity's convergence rate to its minimum value – the Gaussian channel capacity, as $n$ increases.

## III. PROOF OF THEOREM

For the Theorem proof the following time-continuous, additive Gaussian channel is used.

$$\eta(t) = w(t) + \sqrt{q}\int_0^t \xi(s)ds, \quad 0 \le t \le T, \quad (3)$$

where $\xi$ is the channel's input, $E_\xi = \int_0^T \xi^2(t)dt < \infty$, $w$ is a standard Brownian motion independent of $\xi$, and $q$ is the signal-to-noise ratio.

We begin with the following lemma which holds for *any* process $\xi$ that fulfils (3) and Taylor's theorem [7, p95].

**Lemma 1:** As $q \to 0$

$$D(\eta) = \frac{1}{2}D^{(2)}(0)\, q^2 + o(q^2), \quad (4)$$

where $D^{(2)}(0) = \frac{d^2}{dq^2}D(\eta)\bigg|_{q=0}$ and $o(q^2)$ denotes a function $f$ such that $\lim_{q \to 0}[f/q^2] = 0$.

*Proof*

In order to prove the lemma we recall the following relations between the divergence and the minimum mean-square errors [8]:

$$CMMSE(\tilde{\xi}) - CMMSE(\xi) = \frac{2}{q}D(\eta), \quad (5)$$

$$MMSE(\tilde{\xi}) - MMSE(\xi) = 2\frac{d}{dq}D(\eta). \quad (6)$$

In (5), (6) $\tilde{\xi}$ is a Gaussian process with the same covariance function as that of $\xi$,

$$CMMSE(\xi) = E\int_0^T[\{\xi(t) - E[\xi(t)|\eta_0^t]\}^2]dt$$
$$MMSE(\xi) = E\int_0^T[\{\xi(t) - E[\xi(t)|\eta_0^T]\}^2]dt, \quad (7)$$

and $CMMSE(\tilde{\xi})$, $MMSE(\tilde{\xi})$, are defined in a similar way by replacing $\xi$ in (3) and (7) with $\tilde{\xi}$.

It is clear that as $q \to 0$ the values of $CMMSE(\tilde{\xi})$, $CMMSE(\xi)$, $MMSE(\tilde{\xi})$ and $MMSE(\xi)$ exceed $E_\xi$.
Therefore, from (5), (6)

$$D(\eta)_{q=0} = \frac{d}{dq}(D(\eta))_{q=0} = 0. \quad (8)$$

By Taylor's theorem equation (8) implies (4).
Let us choose in (3), the following signal $\xi$,

$$\xi(s) = \frac{1}{\sqrt{n}}\sum_{i=1}^n \sqrt{\frac{2}{T}}X_i \cos\omega_{k_i}s, \quad 0 \le s \le T. \quad (9)$$

In (9) $\omega_{k_i} = \frac{2\pi k_i}{T}$, $i = 1,...,n$ are circular frequencies. We assume in the following that $\omega_{k_i}$ for $i = 1,...,n$, and q are known. Using the above lemma, our first step will be to evaluate the divergence $D(\eta)$. Let us choose the following Fourier family as a complete set of orthonormal functions $\{\phi_i(t)\}$:

$$\frac{1}{\sqrt{T}}, \left\{\sqrt{\frac{2}{T}}\sin\omega_i t\right\}_1^\infty, \left\{\sqrt{\frac{2}{T}}\cos\omega_i t\right\}_1^\infty,$$

where $\omega_i = \frac{2\pi}{T}i$, $i = 1,2,...$.

We represent the Wiener process by

$$w(t) = \underset{n\to\infty}{l.i.m.}\sum_{i=0}^n W_i \int_0^t \phi_i(s)ds,$$

where $W_i, i = 0,1,2,...$ are Gaussian i.i.d. (0, 1). Then, $D(\eta) = D(\eta_1, \eta_2,...)$, where $\eta_i = \int_0^T \eta(t)\phi_i(t)dt$. The only non-Gaussian components that contribute to the divergence are the following

$$\eta_i = \frac{W_i}{\omega_i} - \frac{\sqrt{q/n}}{\omega_i}X_i, \quad i = 1,...,n, \quad (10)$$

where the variables $W_i, i = 1,...,n$ are mutually independent Gaussian random variables with unit variance, independent of $X_i, i = 1,...,n$.

Therefore, the divergence $D(\eta)$ is given in this case by the sum $D(\eta) = \sum_{i=1}^n D(\eta_i)$. Moreover, since divergence is not sensitive to any normalization factor we can choose for each $i$ in (10) $\omega_i = 1$. For the case $n = 1$

$$D_{n=1}(\eta) = D(W_1 + \sqrt{q}X_1).$$

For the $n$-dimensional signal's case (9) we have to replace $q$ in the divergence expression above by $q/n$ and multiply by $n$. Then, by the lemma, for each $q$ (no matter how large, but fixed) and for large $n$

$$D(\eta) = nD(W_1 + \sqrt{\frac{q}{n}}X_1) = n[\frac{1}{2}D^{(2)}(0)\frac{q^2}{n^2} + o(\frac{q^2}{n^2})]$$
$$= \frac{1}{2}D^{(2)}(0)\frac{q^2}{n} + o(\frac{q^2}{n})], \quad n \to \infty. \quad (11)$$

By the data processing law for divergence, for each $n$
$$D(\eta) =$$
$$D(\eta_1,...,\eta_n) \ge D(\sum_{i=1}^n \eta_i) = D(\sum_{i=1}^n W_i + \sqrt{\frac{q}{n}}\sum_{i=1}^n X_i). \quad (12)$$

By normalizing the argument of the divergence in the right hand side of (12) by a factor of $\sqrt{n}$ we get for each $q$

$$D(\eta) \ge D(W + \sqrt{q}\frac{1}{n}\sum_{i=1}^n X_i), \quad (13)$$

Where $W = \frac{1}{\sqrt{n}} \sum_{i=1}^{n} W_i$ is another unit-variance Gaussian random variable, independent of $X_i, i = 1,...,n$.

Observe that $D(W + \sqrt{q} \frac{1}{n} \sum_{i=1}^{n} X_i)$ is a monotonic increasing function of q which equals zero at q=0, and converges to the limit $D(\frac{1}{n} \sum_{i=1}^{n} X_i) = D(\sum_{i=1}^{n} X_i)$ as $q \to \infty$.

For any fixed $q = Q$ we define

$$\Delta_n = D(\sum_{i=1}^{n} X_i) - D(W + \sqrt{Q} \frac{1}{n} \sum_{i=1}^{n} X_i).$$

Since $D(\sum_{i=1}^{n} X_i)$ decreases to zero as $n \to \infty$ (by the convergence in the mean-square sense of $\frac{1}{\sqrt{n}} \sum_{i=1}^{n} X_i$ to a Gaussian random variable),

$$0 \leq \Delta_n \leq D(\sum_{i=1}^{n} X_i) \to 0, \quad n \to \infty.$$

Therefore, $\Delta_n \to 0$ as $n \to \infty$. Let $\varepsilon$ be an arbitrary positive number. Then there exists an integer $N$ such that for all $n \geq N$

$$\Delta_n = D(\sum_{i=1}^{n} X_i) - D(W + \sqrt{Q} \frac{1}{n} \sum_{i=1}^{n} X_i) < \varepsilon. \qquad (14)$$

Inequality (2) follows from (11), (13) and (14).

REFERENCES

[1] A. M. Tulino and S. Verdú, "Monotonic decrease of the non-Gaussianness of the sum of independent random variables: A simple proof," *IEEE Trans. Inform. Theory*, Vol. 52 Issue 9, pp. 4295-4297, Sept. 2006

[2] A. Artstein, K. M. Ball, F. Barthe and A. Naor, "Solution of Shannon's problem on the monotonicity of entropy," *J. American Mathematical Society*, Volume 17, Number 4, pp 975-982, May 12, 2004.

[3] M. Madiman and A. Barron, "The monotonicity of Information in the Central Limit Theorem and Entropy Power Inequalities", *ISIT 2006*, Seattle, USA, pp. 1021-1025, July 2006

[4] A. Barron, "Entropy and the central limit theorem," *Ann. Probab.*, vol. 14, pp. 336–342, 1986.

[5] J. Binia, "On the capacity of certain additive non-Gaussian channels," *IEEE Trans. Inf. Theory*, vol. IT–25, number. 4, pp. 448–452, Jul. 1979.

[6] J. Binia, "New bounds on the capacity of certain infinite dimensional additive non-Gaussian channels," *IEEE Trans. Inf. Theory*, vol. 51, Issue 3, pp. 1218–1221, Mar. 2005.

[7] W. Rudin, *Principles of Mathematical Analysis*, second edition, McGraw-Hill, 1964.

[8] J. Binia, "Divergence and Minimum Mean-Square error in Continuous-Time Additive White Gaussian Noise Channels," *IEEE Trans. Inform. Theory*, Vol. 52 Issue 3, pp. 1160-1163, Mar. 2006.

**Jacob Binia** was born in Jerusalem, Israel, on June 16, 1941. He received his B.Sc., M.Sc., and D.Sc. degrees in electrical engineering from the Technion-Israel Institute of Technology, Haifa, in 1963, 1968, and 1973, respectively. He joined the Armament Development Authority, Ministry of Defense, Israel, in 1963. From 1973 to 1990 he served first as Head of Signal Processing Group and then as Chief Engineer in the Communications Department, RAFAEL – Electronic Division. From 1992 to 1995 he was part of the National Electronic Warfare Research and Simulation Center.

During the years 1985, 1991, while on Sabbaticals from RAFAEL, he was invited by the Electrical Engineering Faculty at the Technion-Israel Institute of Technology to be a Guest Associate Professor.

He is now with New Elective - Engineering Services Ltd., Israel, where he is employed as a communications systems consultant.